\documentclass[fleqn,usenatbib]{mnras}
\usepackage[T1]{fontenc}
\usepackage{ae,aecompl}

\usepackage{graphicx}	% Including figure files
\usepackage{amsmath}	% Advanced maths commands
\usepackage{amssymb}	% Extra maths symbols
\usepackage{xspace}     % adds optional space after abbreviated commands when needed
\usepackage[dvipsnames]{xcolor}
\usepackage[normalem]{ulem}

\newcommand{\kms}{km\:s$^{-1}$\xspace}
\newcommand{\masyr}{mas\:yr$^{-1}$\xspace}
\newcommand{\msun}{$M_\odot$\xspace}

\title[]{Evidence for Two Early Accretion Events That Built the Milky Way Stellar Halo}

\author[Myeong et al.]{
G.~C. Myeong$^1$, E.~Vasiliev$^{1,2}$, G. Iorio$^1$, N.~W. Evans$^1$, V.~Belokurov$^1$ 
\\
$^{1}$Institute of Astronomy, University of Cambridge, Madingley Road, Cambridge CB3 0HA, UK\\
$^2$Lebedev Physical Institute, Leninsky Prospekt 53, Moscow 119991, Russia
}

\pubyear{2019}

\begin{document}
\label{firstpage}
\pagerange{\pageref{firstpage}--\pageref{lastpage}}
\maketitle

\begin{abstract}
The Gaia Sausage is the major accretion event that built the stellar halo of the Milky Way galaxy. Here, we provide dynamical and chemical evidence for a second substantial accretion episode, distinct from the Gaia Sausage. The Sequoia Event provided the bulk of the high energy retrograde stars in the stellar halo, as well as the recently discovered globular cluster FSR 1758. There are up to 6 further globular clusters, including $\omega$~Centauri, as well as many of the retrograde substructures in Myeong et al. (2018), associated with the progenitor dwarf galaxy, named the Sequoia. The stellar mass in the Sequoia galaxy is $\sim 5 \times 10^{7}$ \msun, whilst the total mass is $\sim 10^{10}$ \msun, as judged from abundance matching or from the total sum of the globular cluster mass. Although clearly less massive than the Sausage, the Sequoia has a distinct chemo-dynamical signature. The strongly retrograde Sequoia stars have a typical eccentricity of $\sim0.6$, whereas the Sausage stars have no clear net rotation and move on predominantly radial orbits. On average, the Sequoia stars have lower metallicity by $\sim 0.3$ dex and higher abundance ratios as compared to the Sausage. We conjecture that the Sausage and the Sequoia galaxies may have been associated and accreted at a comparable epoch.
\end{abstract}

\begin{keywords}
Galaxy: stellar content -- Galaxy: halo -- Galaxy: formation -- Galaxy: kinematics and dynamics
\end{keywords}

%%%%%%%%%%%%%%%%%%%%%%%%%%%%%%%%%%%%%%%%%%%%%%%%%%

%%%%%%%%%%%%%%%%% BODY OF PAPER %%%%%%%%%%%%%%%%%%

\section{Introduction}

The Gaia Data Release 2 \citep[DR2,][]{Gaia_Lindegren18} is proving transformational in the identification of substructure in the Milky Way galaxy. This is because substructure retains coherence in phase space over very long timescales~\citep{Jo96,Tr99}. The acquisition of kinematic data, particularly accurate stellar proper motions courtesy of the Gaia satellite, is therefore the key to unlocking the accretion history of the stellar halo. The long-term goal of understanding the building blocks of at least the stellar halo, and perhaps even the entire Galaxy, seems to be within our grasp.

Already, Gaia has provided compelling evidence for the nearly head-on collision of a Magellanic-sized dwarf galaxy with the nascent Milky Way some 8 to 10 billion years ago, the so-called \lq Gaia Sausage'~\citep[see e.g.][]{Belokurov18,Myeong18_ah,Myeong18_sg,Haywood18,Fattahi19}. This name describes the elongated shape of the structure in velocity space. The radial velocity dispersion of the Sausage stars is $\approx 180$ kms$^{-1}$, while the azimuthal and longitudinal dispersions are only $\approx 60$ kms$^{-1}$ \citep[see e.g., Figure 4 of][]{Belokurov18}. The name therefore follows the long-standing scientific practice of being descriptive and informative\footnote{EV is unhappy with the name and is looking for a better one.}.

The aftermath of this accretion event is detectable in the inner stellar halo of the Galaxy as a giant cloud of relatively metal-rich ([Fe/H]$\gtrsim-1.5$) stars on highly radial orbits. Originally traced with nearby Main Sequence stars \citep{Belokurov18, Myeong18_ah}, the Sausage debris has now been found over a large distance range with a number of distinct tracers, including Blue Horizontal Branch stars \citep{Deason18, Lancaster18} and RR Lyrae \citep{Simion19,Iorio19}. The characteristic property of the residue of this collision is that the orbits are eccentric with little or no net angular momentum. The debris of this event does not provide any strongly prograde or retrograde material, as befits an almost head-on collision.

Alternatively, it was proposed that an ancient major merger -- dubbed \lq Gaia-Enceladus' -- could have given rise to the bulk of the retrograde stars in the halo, as well as some of the low-angular momentum debris \citep{He18}. The fundamental difference between the two hypotheses is that the \lq Gaia-Enceladus' encompasses not just the highly eccentric component of the halo, but also the strongly retrogarde component. For example, in \citet{He18}, the `Gaia-Enceladus' stars have angular momentum component satisfying $-15000< J_\phi < 150$ kms$^{-1}$ kpc$^{-1}$, independent to the total energy, and so span a range from mild prograde through highly eccentric to strongly retrograde. The question of whether a single collision could produce such a spray of debris with different kinematical properties remains open.

In fact, the suggestion that the retrograde component of the halo may have been accreted already predates the arrival of the Gaia data by many years~\citep[see e.g.,][]{Norris89, Carollo07, Beers12, Majewski12}. The retrograde and peculiar globular cluster $\omega$~Centauri has also long been suspected of playing a role in the supply of retrograde stars, as it may be the stripped nucleus of a dwarf galaxy~\citep{Be03}. The Gaia data releases have provided new samples of the retrograde halo component, which have been scoured for evidence of multiple minor mergers and accretion events~\citep[e.g.,][]{Helmi17, Myeong18_ah, Myeong18_rg}. The question therefore  at issue is: did one merger event provide both the eccentric and retrograde components of the stellar halo (as in the \lq Gaia-Enceladus' theory) or does the retrograde component have a different origin from the eccentric component (the \lq Gaia-Sausage' theory)?

We provide a possible answer to this question in our paper, but our line of reasoning begins in a roundabout way with another unusual retrograde object,  FSR~1758. This was  originally discovered by \citet{Froebrich2007} as a claimed open cluster and later identified as a globular cluster \citep{Cantat2018}. \citet{Barba19} recently reported the first estimate of its distance and noticed its unusual size, using a combination of data from the DECam Plane Survey (DECaPS, \citealt{Schlafly18}) and the VISTA Variables in the Via Lactea (VVV) Extended Survey, complemented with Gaia DR2. It is an extended agglomeration of stars, located at ($\ell = 349^\circ, b = 3^\circ$) and with a heliocentric distance of $10-12$~kpc. They determined the core radius of FSR~1758 to be $\approx 10$ pc, and estimated the tidal radius to be $\approx 150$ pc. Considering its unusual size, \citet{Barba19} questioned whether FSR~1758 is the remnant of a dwarf galaxy or an unusually large globular cluster. Subsequently, \cite{Simpson19} found 3 stars in the centre of FSR~1758 with line-of-sight velocities from the Gaia Radial Velocity Spectrograph (RVS) and argued that the object has a line-of-sight velocity of $227 \pm 1$ kms$^{-1}$. Although the 3 stars are insufficient to come to a definite conclusion as regards FSR~1758's internal velocity dispersion, nonetheless \citet{Simpson19} argued on the basis of its highly retrograde orbit that it is an accreted halo globular cluster.

In fact, the globular cluster (GC) datasets have been scrutinised carefully for evidence of accretion events in recent years. They have proved surprisingly powerful tracers of the merger events that build the stellar halo of the Galaxy. This was made clear in \citet{Forbes10}, who showed that the globular clusters that formerly belonged to the Sagittarius galaxy follow a different track in the plane of age-metallicity as compared to the bulk of the primordial or in situ clusters. Subsequently, \citet{Myeong18_sg} identified a sample of 10 high eccentricity, high-energy, old halo GCs strongly clumped in action space that belonged to the `Gaia Sausage' event. More speculatively, \citet{Kr19} used the age--metallicity distribution of Galactic globular clusters to reconstruct the entire merger history, claiming three substantial events. If FSR~1758 is indeed an accreted GC, then this suggests that a systematic search for companion GCs accreted in the same event, as well as other stellar debris such as substructures and tidal tails, may provide a picture of the progenitor. 

In Section 2, we use Gaia's kinematic data in combination with photometry from DECaPS to determine structural parameters and the proper motion dispersion profile for FSR~1758, and show that its declining fall-off is characteristic of a GC. The nature of FSR~1758 having been established, we search for companion GCs and stellar substructures moving on orbits of similar eccentricity and inclination that may have joined the Milky Way in the same accretion event in Section 3. \citet{Barba19} introduced the picturesque term Sequoia to describe the size of FSR~1758. We retain the term and slightly adapt it for our own use. In our picture, FSR~1758 is one of about five or more GCs that populated the Sequoia dwarf galaxy, whose existence was already conjectured from our stellar substructure searches~\citep{Myeong18_ah,Myeong18_rg}. Its disruption brought these GCs into the Milky Way on similar orbits, as well as abundant retrograde high energy stellar substructure. We argue that the remnants of the Sequoia galaxy are dynamically distinct from the Gaia Sausage as they are retrograde, whereas the Sausage was an almost head-on collision. The dual pattern of these accretion events is evident in energy and actions, and is also shown clearly when the chemical evidence is analysed. The stars and substructures associated with the Sequoia have different mean metallicities and different abundance ratios. In Section 4, we provide estimates of the age and mass of the Sequoia galaxy and compare with the Gaia Sausage. Finally, we summarise our results in our concluding Section 5.

\section{The Nature of FSR~1758}

\begin{figure*}
\includegraphics{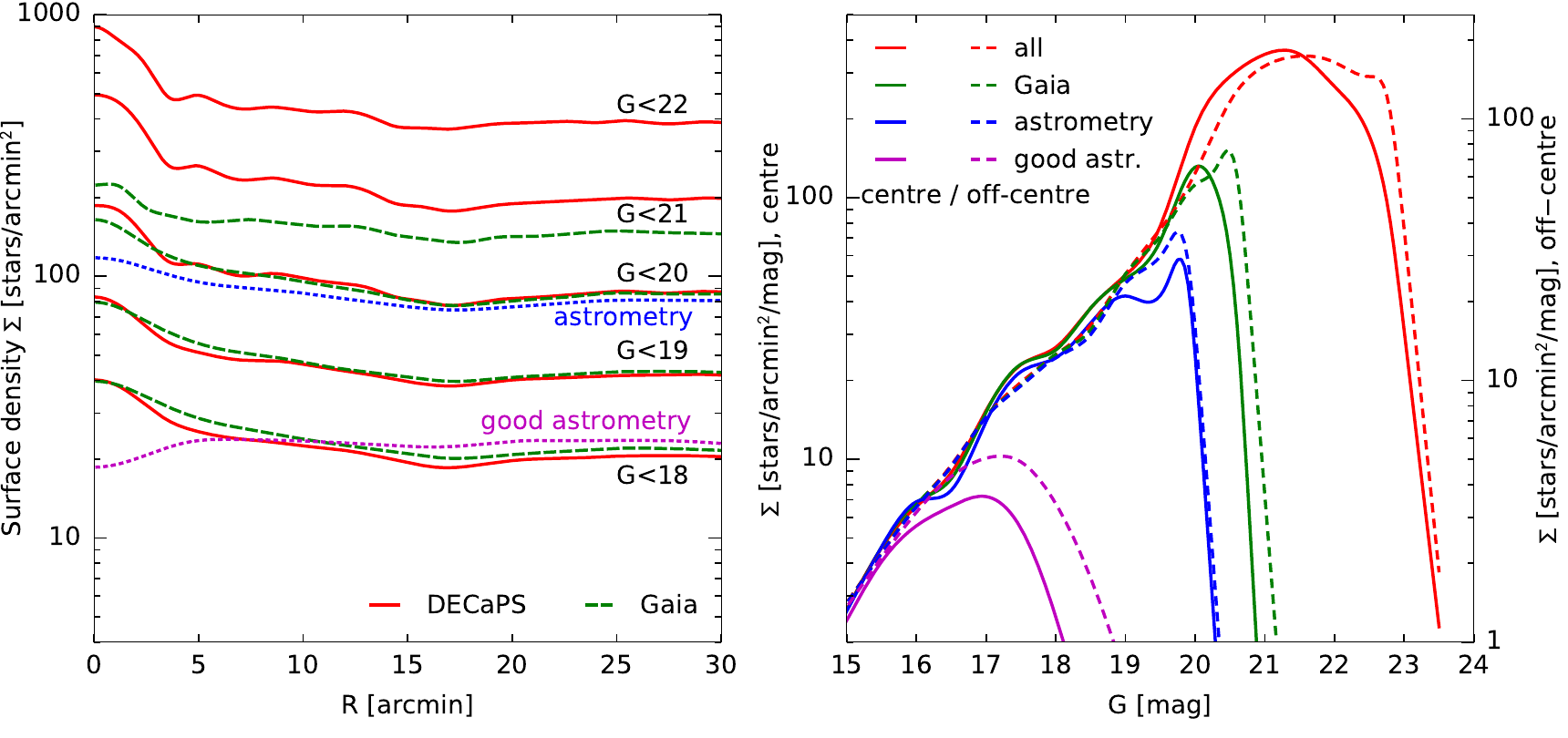}
\caption{Left: Radial profiles of the surface density of FSR~1758. Solid red curves show the DECaPS data and dashed green the Gaia data; the former have been converted to Gaia $G$ magnitude using eqn~\ref{eq:Gproxy}. If the magnitude distribution of stars were not spatially varying, these curves would have a constant vertical offset (in logarithmic units). It is clear that the Gaia data is reasonably complete up to $G\le 19$ in the centre and up to $G\le 20$ elsewhere, while the DECaPS data is fairly complete up to $G\approx 21$. Right: Distribution of stars as a function of magnitude in the central 2 arcmin (solid lines) and in an off-centered field at $R\sim 11$ arcmin to the north (dashed lines); the latter is vertically offset by a factor of 2 to compensate for the lower overall density of stars. It illustrates the same points about the completeness of Gaia data as compared to DECaPS (the difference between red and green lines starts to appear at $G\gtrsim 20$). In the central area, there is an excess of stars with $20\le G\le 21$. These are numerous main-sequence stars of the cluster, absent in off-centered fields. Therefore, we need to include the stars up to $G\le 21$ in order to have a faithful representation of surface density of the cluster.
} \label{fig:completeness}
\end{figure*}

\subsection{Data}

First, we cross-match the positions of stars between Gaia and DECaPS, using a search radius of 0.5''. DECaPS provides roughly two magnitudes deeper photometry than Gaia, but it is saturated for bright stars and has patchy spatial coverage, so we study the union of the two datasets. We use the following combinations of $r$ and $i$ DECaPS photometric bands as a proxy for Gaia $G$ and $G_\mathrm{BP}-G_\mathrm{RP}$ (derived by comparing the magnitudes of cross-matched stars):
\begin{equation}  \label{eq:Gproxy}
\begin{aligned}
G &\approx r + 0.1 + 0.3 (r-i) - 0.5 (r-i)^2 ,\\
G_\mathrm{BP}-G_\mathrm{RP} &\approx \;0.65\; +\, 2.35 (r-i) - 0.3 (r-i)^2 .
\end{aligned}
\end{equation}
Fig.~\ref{fig:completeness} shows the surface density profiles of stars in different ranges of magnitudes, and their distributions by magnitudes at different spatial locations. By comparing the density of stars in Gaia and DECaPS datasets, we conclude that the former is reasonably complete up to $G\lesssim 20$. The subset of stars with astrometric measurements also extends roughly to $G=20$, but is less complete in the central area. We wish to include the fainter stars without astrometry in order to mitigate the bias in representation of the spatial density profile of the cluster. We chose to use stars up to $G=21$, of which roughly 70\% are present in the Gaia dataset, and only 40\% have astrometric measurements. During fitting, we also infer the total mass from the intrinsic (error-deconvolved) proper motion (PM) dispersion. We only use a high-quality subset of stars (marked as ``good astrometry'' in the figure) for this inference, ignoring all sources with \texttt{astrometric\_excess\_noise}${} > 1$ or \texttt{phot\_bp\_rp\_excess\_factor}${} > 1.3 + 0.06(G_\mathrm{BP}-G_\mathrm{RP})^2$, as suggested by \citet{Gaia_Lindegren18}. 

%%%%%%%%%%%%%%%%%%%%%%%%%%%%%%%%%%%%%%%%%%%%%%%%%%%%%%%%%%%%%
\subsection{Dynamical Modelling}

We use a probabilistic model, in which the stars are drawn from a mixture of two populations: the cluster and the field. The distribution of field stars is assumed to be spatially uniform, and described by a sum of two bivariate Gaussians in the PM plane. We assume that the density of cluster stars follows a generalized King profile, also known as the \textsc{limepy} family of models \citep{Gieles15}, which has been shown to adequately describe realistic globular clusters \citep{HenaultBrunet19}. It has the following free parameters: mass $M$, scale (core) radius $R_c$, dimensionless potential depth at the centre (King parameter) $W_0$, and truncation parameter $g$ controlling the density profile in the outer parts. For the models in this section, we do not assume any particular relation between the total cluster mass and the number of observed cluster members $N^\mathrm{clust}$ (this relation is examined in the next section). Rather, the total mass of the cluster manifests itself only kinematically, through the overall amplitude of velocity dispersion. We assume a Gaussian distribution for the PM of cluster stars, centred around its mean PM, and with a spatially-variable width. 

We measure the parallax distribution of field stars directly from the data using stars outside the central 5 arcmin and represent it by a mixture of three Gaussian components, with parameters fixed throughout the rest of the modelling. The intrinsic (error-free) parallaxes of cluster stars are assumed to be equal to the inverse distance to the cluster (fixed to $D=10$~kpc), plus the constant zero-point parallax offset $-0.03$~mas \citep{Gaia_Lindegren18}. We do not use any information about colours and magnitudes for membership determination, because they are severely affected by spatially-variable reddening.

\begin{figure*}
\includegraphics{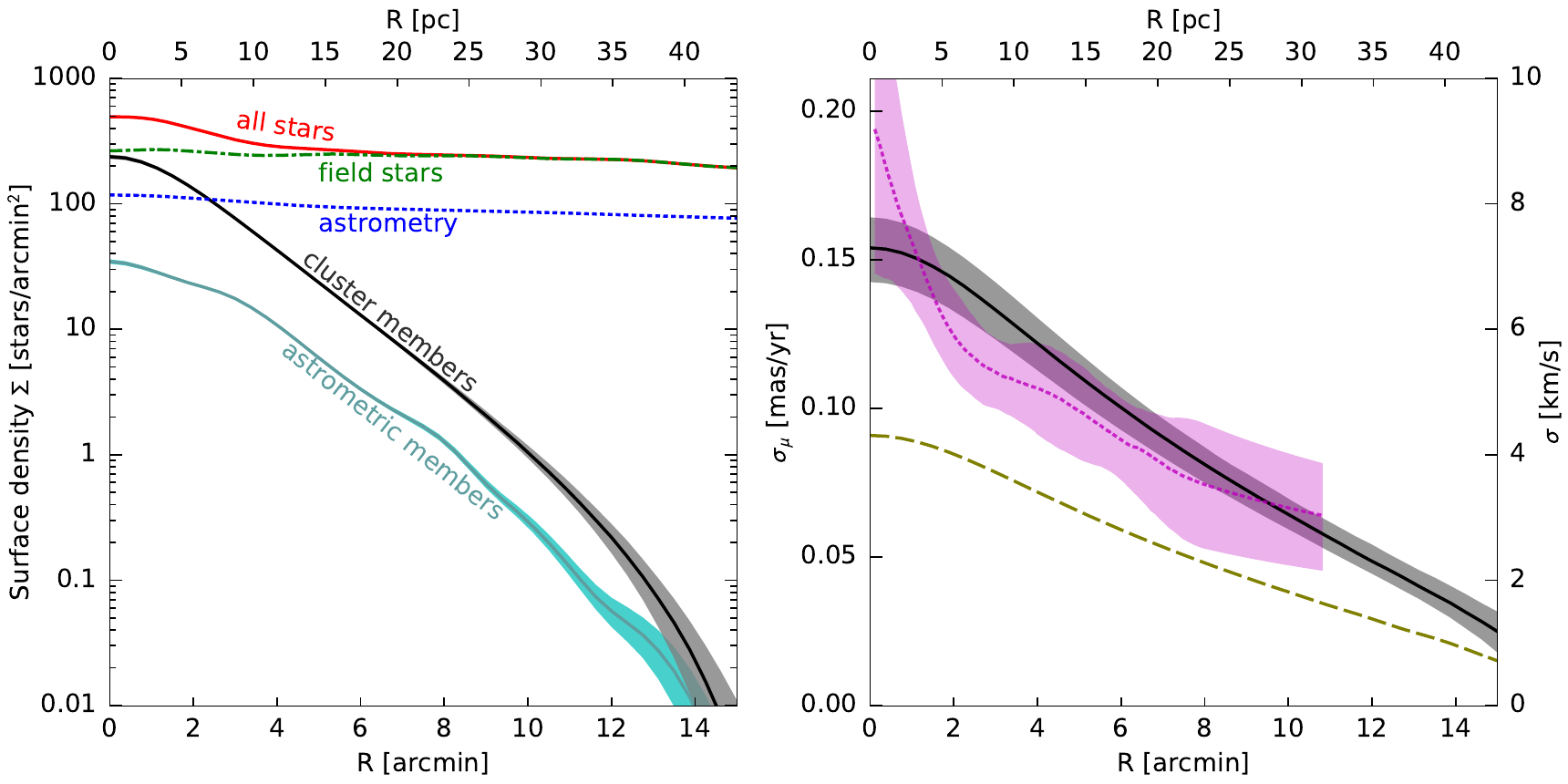}
\caption{Left: Surface density profiles for different subsets of stars. Red solid line: all stars with $G<21$ (same as the corresponding curve in Fig.~\ref{fig:completeness}). Black solid line and gray shaded region: all cluster members. Green dot-dashed line: field stars (the difference between the previous two curves). Blue dotted line: all stars with astrometric measurements. Cyan shaded region: cluster stars with astrometric measurements. Right: PM dispersion profile of cluster stars. Black line and shaded region: parametric profile of the generalized King model, with 68\% confidence interval. Purple dotted line and shaded region: non-parametric estimate from stars with $\gtrsim 80\%$ membership probability. Olive dashed line: the profile of a King model with the total mass $2.5\times10^5\,M_\odot$, as inferred by the photometric fit (Fig.~\ref{fig:cmd}).
} \label{fig:profiles}
\end{figure*}
%%%%%%%%%%%%%

The likelihood of observing a star $i$ at a given distance from the cluster centre, with or without further astrometric information, is given by a sum of likelihoods of the two alternative hypotheses:
\begin{equation}
\mathcal L_i = \mathcal L_i^\mathrm{clust} + \mathcal L_i^\mathrm{field}.
\end{equation}
Let $N^\mathrm{clust}$ be the (unknown) total number of cluster members in our sample of stars within the radius $R_\mathrm{max}$. We choose $R_\mathrm{max} \gg R_c$, so that all possible cluster members are included, and normalize the cluster surface density $\Sigma^\mathrm{clust}(R)$ so that
$\int_0^{R_\mathrm{max}} \Sigma^\mathrm{clust}(R)\;2\pi\,R\,\mathrm dR = N^\mathrm{clust}$. 
The remaining $N^\mathrm{field} \equiv N^\mathrm{total}-N^\mathrm{clust}$ observed stars are then attributed to the field population with a spatially-uniform density 
$\Sigma^\mathrm{field} \equiv N^\mathrm{field} / (\pi\,R_\mathrm{max}^2)$.
Then the likelihoods of a given star to belong to either population are
\begin{equation}
\mathcal L_i^\mathrm{clust} \equiv \Sigma^\mathrm{clust}(R_i) \;A^\mathrm{clust}_i, \qquad
\mathcal L_i^\mathrm{field} \equiv \Sigma^\mathrm{field} \;A^\mathrm{field}_i,
\end{equation}
where the factors $A^\mathrm{clust}_i,\: A^\mathrm{field}_i$ are unity for stars without astrometric measurements, or describe the likelihood of measuring the observed parallax and PM, given the intrinsic distribution functions of either population, convolved with measurement uncertainties. For the cluster population, this factor is a product of the parallax likelihood and the PM likelihood (ignoring correlations between them):
\begin{equation}
\begin{aligned}
A^\mathrm{clust}_i &\equiv 
\mathcal N(\varpi_i-\varpi^\mathrm{clust},\; \epsilon_{\varpi,i}^2) \;\;
\mathcal N(\boldsymbol\mu_i-\boldsymbol\mu^\mathrm{clust},\; \mathsf S_i) , \\
\mathsf S_i &\equiv \begin{pmatrix}
\epsilon_{\mu_\alpha,i}^2 + \sigma_\mu^2(R_i)  &
\rho_i\:\epsilon_{\mu_\alpha,i}\:\epsilon_{\mu_\delta,i}  \\
\rho_i\:\epsilon_{\mu_\alpha,i}\:\epsilon_{\mu_\delta,i}  &
\epsilon_{\mu_\delta,i}^2 + \sigma_\mu^2(R_i)
\end{pmatrix} ,
\end{aligned}
\end{equation}
where $\mathcal N$ is the uni- or bivariate normal distribution, $\varpi$ is the parallax, $\boldsymbol\mu\equiv\{\mu_\alpha,\mu_\delta\}$ is the PM with associated measurement uncertainties $\epsilon$, $\rho_i$ is the correlation coefficient between the two components of PM uncertainty matrix, $R_i$ is the distance of the star from the cluster centre, $\sigma_\mu(R)$ is the spatially-dependent intrinsic PM dispersion of cluster stars, whose amplitude is proportional to the square root of the cluster mass. We use only a subset of stars with reliable PM measurements outside the central 2 arcmin for inferring the intrinsic dispersion, since the stars in the centre may be affected by crowding. The PM of remaining (mostly faint) stars are still used to determine the membership, but in doing so, we use a conservative value $\sigma_\mu=0$. For the field population, the expressions are similar, but involve several Gaussian components (two for PM and three for parallax), with the intrinsic PM dispersion being a spatially-constant symmetric $2\times2$ matrix rather than a single spatially-varying quantity.

The fitting procedure optimizes the model parameters ($N^\mathrm{clust}, \boldsymbol\mu^\mathrm{clust}, \boldsymbol\mu^\mathrm{field}$, covariance matrices of the field population, parameters of the cluster density profile) to maximize the total log-likelihood
\begin{equation}
\ln\mathcal L \equiv \sum_{i=1}^{N^\mathrm{total}} \ln\mathcal L_i ,
\end{equation}
via Markov Chain Monte Carlo as implemented in the \textsc{emcee} package \citep{ForemanMackey13}. We then evaluate the posterior probability of membership for each star as
\begin{equation}
P^\mathrm{clust}_i = \frac{\mathcal L^\mathrm{clust}_i}
{\mathcal L^\mathrm{clust}_i + \mathcal L^\mathrm{field}_i} .
\end{equation}
The total number of cluster members is $N^\mathrm{clust} = \sum_i P^\mathrm{clust}_i$. We stress that we do not make hard cuts in any of the observed quantities (parallax, PM, radius) to separate the cluster and the field populations. For stars with astrometric measurements, the membership probability distribution is strongly bimodal (the two populations are well separated), while for stars without astrometry, the membership probability smoothly drops with radius from $\sim 0.5$ in the centre down to zero at large radii. In total, we have $N^\mathrm{clust}\simeq 7500$ member stars, of which only $\sim 1600$ are astrometrically selected, and only $\sim 350$ are used in the dynamical mass determination through the intrinsic PM dispersion.   

\begin{figure*}
\includegraphics{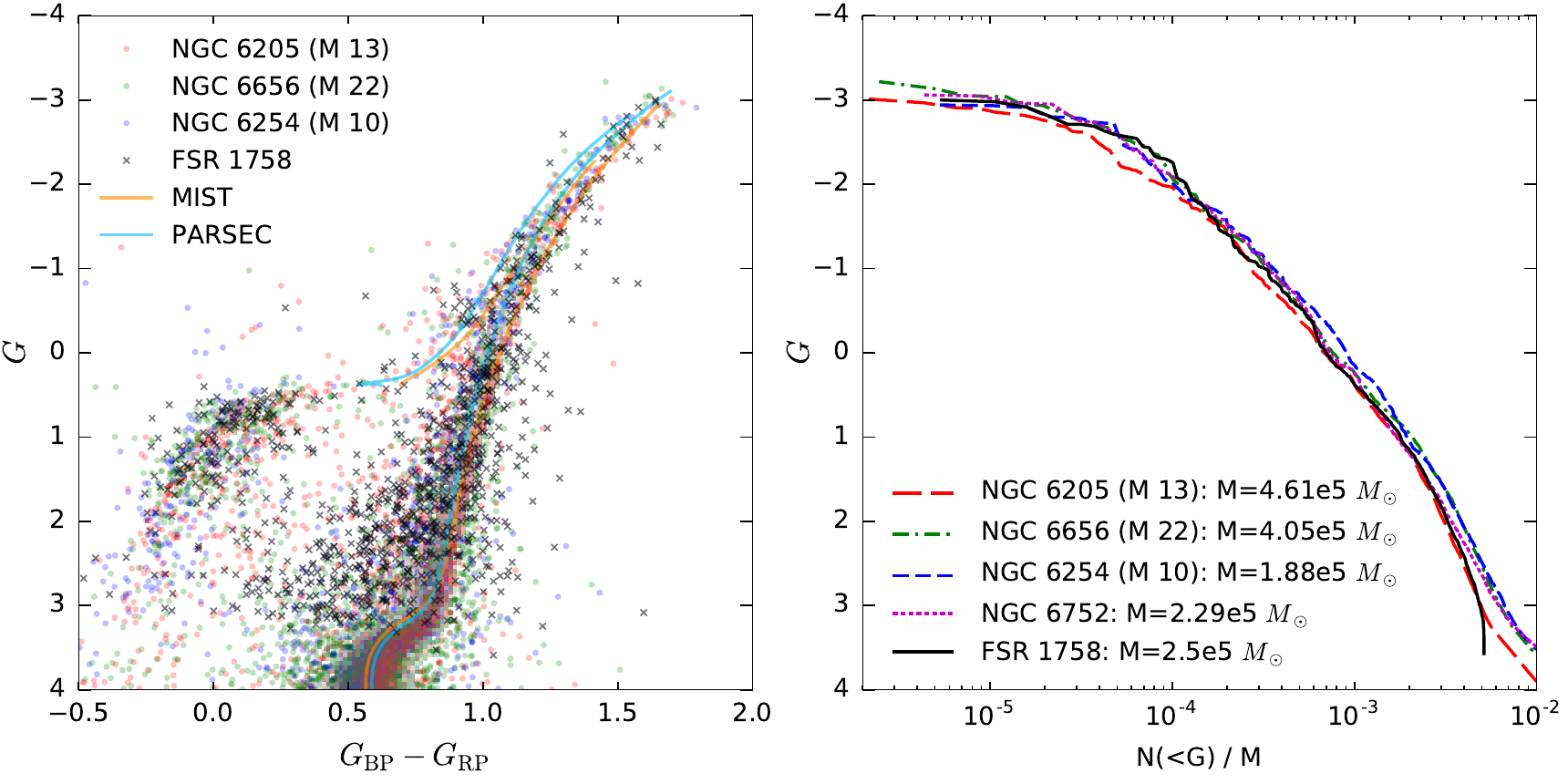}
\caption{Left: De-reddened colour--absolute magnitude diagrams (CMD) of several Galactic globular clusters with similar metallicity [Fe/H]$\approx-1.5$, shown by coloured dots, together with the one for FSR~1758, shown by black crosses. We used the distance $D=10$~kpc and reddening $E(B-V)=0.73$ for the latter cluster, and the literature values for the remaining ones. Overplotted are theoretical isochrones for 12.5~Gyr old population from two stellar-evolution models: MIST \citep{Choi16} and PARSEC \citep{Bressan12}. Right: Cumulative number of stars brighter than the given magnitude (horizontal axis) as a function of magnitude (vertical axis), scaled to the mass of each cluster. The masses of 4 NGC clusters are from \citet{Baumgardt19}, while the mass of FSR~1758 is inferred as the best match to the remaining clusters.
} \label{fig:cmd}
\end{figure*}

Fig.~\ref{fig:profiles} shows the inferred cluster density and the PM dispersion profiles. The best-fit parameters of the generalized King model are: King radius of 3~arcmin ($\sim 8.5$~pc for the assumed distance $D=10$~kpc), King parameter $W_0\simeq 3$, truncation radius $\sim 15$~arcmin, and truncation parameter $g\simeq 1$. The core radius (defined as the projected distance from the cluster centre where the surface density drops to 1/2 its central value) is $\sim 6.5$~pc, and the half-light radius (the projected distance enclosing half of the cluster stars) is $\sim 9$~pc. The core radius is somewhat smaller than determined by \citet{Barba19} from their fit to the DECaPS photometric sample (without considering astrometry). On the other hand, the truncation radius ($\sim 50$~pc) is three times smaller than found in that paper. We stress that the density profile in the outer parts, and in particular the truncation radius, is determined mainly by Gaia astrometry, so is more reliably constrained than just using the photometry alone.

Overall, the modelling procedure makes good use of both the deeper DECaPS photometry in the cluster centre and the Gaia astrometry in the outer parts. However, the inferred cluster mass (equivalently, the intrinsic PM dispersion) appears to be rather high, $M\simeq (7\pm1)\times10^5\,M_\odot$, compared to the photometric model of the next section. We stress that the width of the intrinsic PM distribution is inferred by convolving it with the measurement uncertainties and comparing the error-broadened distribution with the actually observed one. Hence, it strongly depends on the reliability of uncertainty estimates $\epsilon_{\mu}$ of stars in the Gaia dataset. Even for the high-quality subsample, these errors are in the range $0.1-0.3$~\masyr, comparable to or exceeding the inferred value of intrinsic PM dispersion. It is known that the formal uncertainties in Gaia PMs are underestimated \citep[e.g.,][]{Gaia_Lindegren18}. We multiplied the uncertainties quoted in the catalogue by a correction factor 1.1, as suggested in that paper, before running the fit. If we instead increase the uncertainties by a factor 1.3, this reduces the PM dispersion by a third, bringing it into agreement with the total cluster mass estimated from photometry.

Finally, to check if the PM dispersion profile could be biased by the assumed parametric form of the generalized King model, we also determined it non-parametrically from the subset of high-quality stars classified as cluster members with $\ge 80\%$ probability. We used the method of \citet{Vasiliev18}, representing $\sigma_\mu(R)$ as a cubic spline with the values at four control points adjustable during the fit, while taking into account spatially correlated measurement errors. The result, shown in the right panel of Fig.~\ref{fig:profiles}, agrees reasonably well with the parametric profile, but is higher in the very centre. This is likely caused by crowding issues, and for this reason we have excluded the stars in the central two arcmin from the high-quality sample used to determine the PM dispersion in the parametric fit. In any case, the PM dispersion profile appears to be declining with radius, which is natural to expect for a globular cluster, but not for a dwarf galaxy.

\begin{figure}
\includegraphics{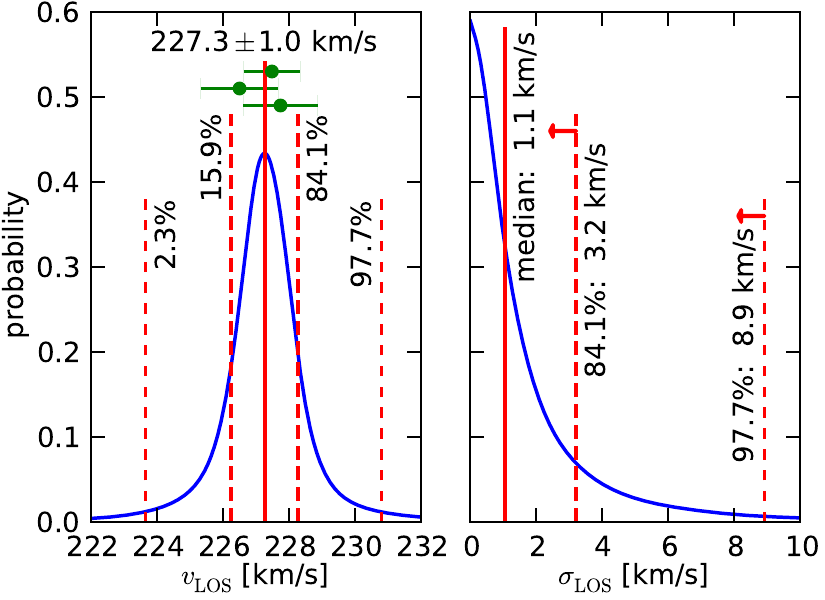}
\caption{Estimate of the mean line-of-sight velocity (left panel) and its dispersion (right panel) from the three stars detected by the Gaia RVS instrument. The measured values, shown by green dots with error bars, are very similar and consistent with zero intrinsic dispersion, however, the probability distributions as shown in this figure are heavy-tailed, and the chances of $\sigma$ exceeding 5~\kms are around 10\%.
} \label{fig:lineofsightvelocity}
\end{figure}

%%%%%%%%%%%%%%%%%%%%%%%%%%%%%%%%%%
\subsection{Photometric modelling}

Since almost all stars with magnitudes $G\lesssim 19-20$ have astrometric measurements and are well separated into the cluster and the field populations, we may use the sample of astrometrically confirmed members to determine the total cluster mass. The idea is to compare the distribution of stars by magnitudes with those of several other clusters with similar colour-magnitude diagrams (CMDs), for which the total mass and the distance are known. By normalizing the number of stars brighter than the given absolute magnitude by the cluster mass, we constructed the cumulative magnitude distribution profiles for a dozen globular clusters in the range of metallicities $-2\lesssim \mathrm{[Fe/H]} \lesssim -1$. The masses and distances of these clusters are taken from \citet{Baumgardt19}, who used a large library of $N$-body simulations and a variety of observational constraints to measure the masses. Three clusters have very similar CMDs to FSR~1758, particularly with regards to the location of blue horizontal branch (BHB): NGC~6205 (M~13), NGC~6254 (M~10) and NGC~6656 (M~22). They all have metallicities $\simeq -1.5$ and masses $(2-5)\times 10^5\,M_\odot$. 

Fig.~\ref{fig:cmd}, left panel, shows the composite CMDs of these three clusters, with the astrometrically detected members of FSR~1758 overplotted by black crosses. The measured colour $G_\mathrm{BP}-G_\mathrm{RP}$ and the $G$-band magnitude are de-reddened using the coefficients given in Table~2 of \citet{Babusiaux18}, and the observed magnitude is converted to the absolute magnitude. By matching the stars of FSR~1758 to the empirically determined isochrones, we infer the reddening coefficient $E(B-V)=0.73$ and the distance $D=10$~kpc, in reasonable agreement with \citet{Barba19}. Stars in the lower part of the red giant branch (RGB) are systematically offset to the left from the isochrone, but this is expected for such a dense and highly reddened region. If we use DECaPS $r$ and $i$ bands instead of Gaia $G_\mathrm{BP}-G_\mathrm{RP}$, as per eqn~\ref{eq:Gproxy}, the scatter and offset of member stars from the isochrone curve at faint magnitudes are substantially reduced.  The upper part of the RGB and the BHB of FSR~1758 match well the location of these features in the other three clusters.

The right panel of Fig.~\ref{fig:cmd} shows the cumulative number of stars as a function of magnitude, normalized by the mass of each cluster. The three clusters listed above have similar profiles, and in fact almost all other clusters also follow the same trend (we show additionally the cluster NGC~6752, which has a somewhat higher metallicity). By matching, we infer the mass of FSR~1758 to be $\sim 2.5\times 10^5\,M_\odot$, with an uncertainty $\lesssim 20\%$. 

As noticed by \citet{Simpson19}, three bright stars in the centre of FSR~1758 are actually present in the Gaia RVS sample, having values of line-of-sight velocity around 227~\kms. 
\citet{Simpson19} additionally reported a fourth star with a very similar line-of-sight velocity at a projected distance $\sim0.6^\circ$ from the cluster. Given that its PM, $G$ magnitude and colour are all very close to those of the three other stars, it is unlikely to be a field star, but it also cannot be a current cluster member, being more than twice as far as the inferred cutoff radius. This star may have been tidally stripped from the cluster.

With only three stars, it is not possible to put strong constraints on the internal velocity dispersion $\sigma$. The measured values are very close to each other and consistent with zero intrinsic scatter, but values of $\sigma$ up to 5--10~\kms are not strongly excluded (Fig~\ref{fig:lineofsightvelocity}). 
These values are also consistent with the photometrically estimated mass, which corresponds to the central velocity dispersion $\sim 4$~\kms, and even with the higher values inferred from PM, although these are less reliable.

 \begin{table*}

 \caption{The kinematic, action, and orbital properties of the probable and possible GCs from Monte Carlo sampling.}

 \begin{center}
   \begin{tabular}{lrrrrrrrc}
 \hline \hline
 \multicolumn{1}{c}{Name} &
 \multicolumn{1}{c}{$(J_R,J_\phi,J_z)$} &
 \multicolumn{1}{c}{E} &
 \multicolumn{1}{c}{pericentre} &
 \multicolumn{1}{c}{apocentre} &
 \multicolumn{1}{c}{ecc.} &
 \multicolumn{1}{c}{incl.} \\

 \multicolumn{1}{c}\null &
 \multicolumn{1}{c}{(kpc~\kms)} &
 \multicolumn{1}{c}{(km$^2$s$^{-2}$)} &
 \multicolumn{1}{c}{(kpc)} &
 \multicolumn{1}{c}{(kpc)} &
 \multicolumn{1}{c}{\null} &
 \multicolumn{1}{c}{(deg)}  \\

\hline

FSR~1758 & ($620^{+160}_{-130},-1250^{+150}_{-150},230^{+50}_{-40}$) & $-142500^{+8100}_{-8400}$ & $3.8^{+0.5}_{-0.5}$ & $16.2^{+3.0}_{-2.6}$ & $0.62^{+0.02}_{-0.01}$ & $146.5^{+1.4}_{-1.4}$ \\ 
\hline 
NGC~3201 & ($900^{+140}_{-120},-2800^{+80}_{-90},310^{+20}_{-20}$) & $-112300^{+3200}_{-3000}$ & $8.4^{+0.1}_{-0.1}$ & $29.3^{+2.6}_{-2.1}$ & $0.55^{+0.03}_{-0.02}$ & $152.6^{+0.3}_{-0.3}$ \\ 
\hline 
$\mathrm{\omega}$~Centauri & ($270^{+30}_{-30},-520^{+30}_{-30},100^{+20}_{-20}$) & $-185000^{+900}_{-800}$ & $1.5^{+0.1}_{-0.1}$ & $7.2^{+0.1}_{-0.2}$ & $0.65^{+0.02}_{-0.03}$ & $139.7^{+2.2}_{-2.1}$ \\ 
\hline 
NGC~6101 & ($1400^{+240}_{-210},-3210^{+200}_{-210},800^{+60}_{-50}$) & $-97200^{+4400}_{-4400}$ & $10.7^{+0.5}_{-0.5}$ & $41.4^{+4.7}_{-4.2}$ & $0.59^{+0.02}_{-0.02}$ & $143.1^{+0.5}_{-0.5}$ \\ 
\hline 
NGC~6535 & ($140^{+30}_{-30},-350^{+30}_{-30},66^{+8}_{-7}$) & $-207600^{+800}_{-700}$ & $1.3^{+0.1}_{-0.1}$ & $4.5^{+0.2}_{-0.1}$ & $0.56^{+0.05}_{-0.04}$ & $159.0^{+2.2}_{-2.4}$ \\ 
\hline 
\hline 
NGC~6388 & ($80^{+10}_{-5},-250^{+40}_{-30},71^{+1.4}_{-1.3}$) & $-222300^{+3800}_{-3300}$ & $0.9^{+0.1}_{-0.1}$ & $3.5^{+0.2}_{-0.5}$ & $0.59^{+0.03}_{-0.02}$ & $149.2^{+3.2}_{-4.3}$ \\ 
\hline 
NGC~6401 & ($54^{+10}_{-7},-590^{+110}_{-110},150^{+30}_{-30}$) & $-194000^{+8200}_{-8900}$ & $2.5^{+0.5}_{-0.5}$ & $4.9^{+0.9}_{-0.8}$ & $0.34^{+0.02}_{-0.02}$ & $142.5^{+0.9}_{-1.1}$ \\ 
\hline 

   \end{tabular}
 \end{center}
  \label{tab:table_1}
 \end{table*}

\begin{figure*}
 \begin{center}
\includegraphics[width=0.7\textwidth,]{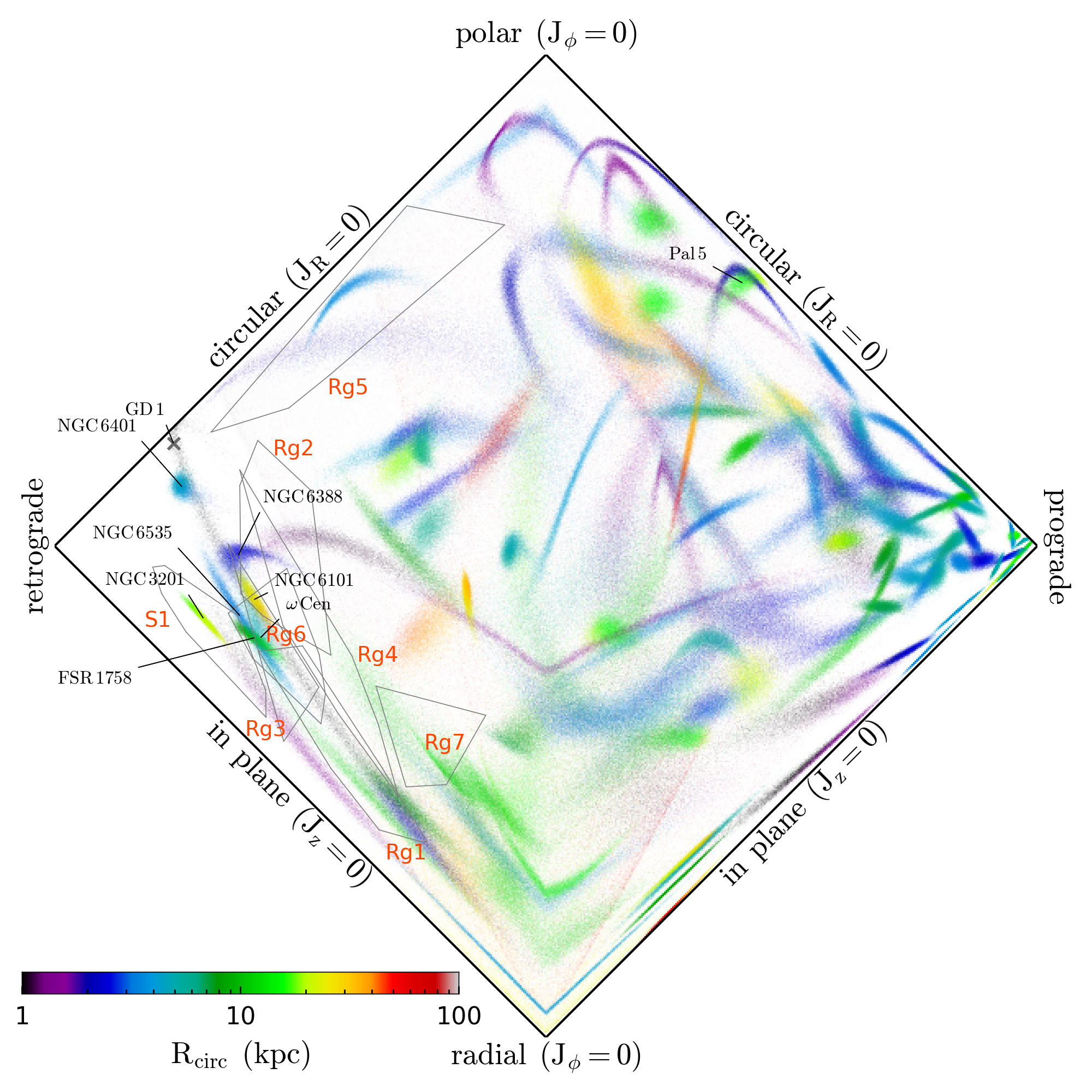}
 \end{center}
 \caption{The action-space map for the Milky Way GCs \citep{Vasiliev19} and retrograde substructures \citep{Myeong18_rg}. The GD-1 stream \citep{Grillmair06} is also marked with a cross based on a representative 6D phase space information from \citet{Webb19}. The horizontal axis is ($J_{\phi} / J_{\mathrm{tot}}$), and the vertical axis is $(J_z - J_R) / J_{\mathrm{tot}}$), analogous to Fig.~5 of \citet{Vasiliev19}. Colour marks the circular orbit radius for the corresponding total energy ($R_{\mathrm{circ}}(E_\mathrm{tot})$). Each object is shown with 1000 Monte Carlo representations of the orbit as drawn from the observational errors. The geometry of the figure can be thought as a projection of the energy-scaled three-dimensional action-space, viewed from the top~(cf. Fig. 3.25 of \citealt{BT08}).}
 \label{fig:actionmap}
 \end{figure*}
\begin{figure}
 \begin{center}
\includegraphics[width=0.45\textwidth]{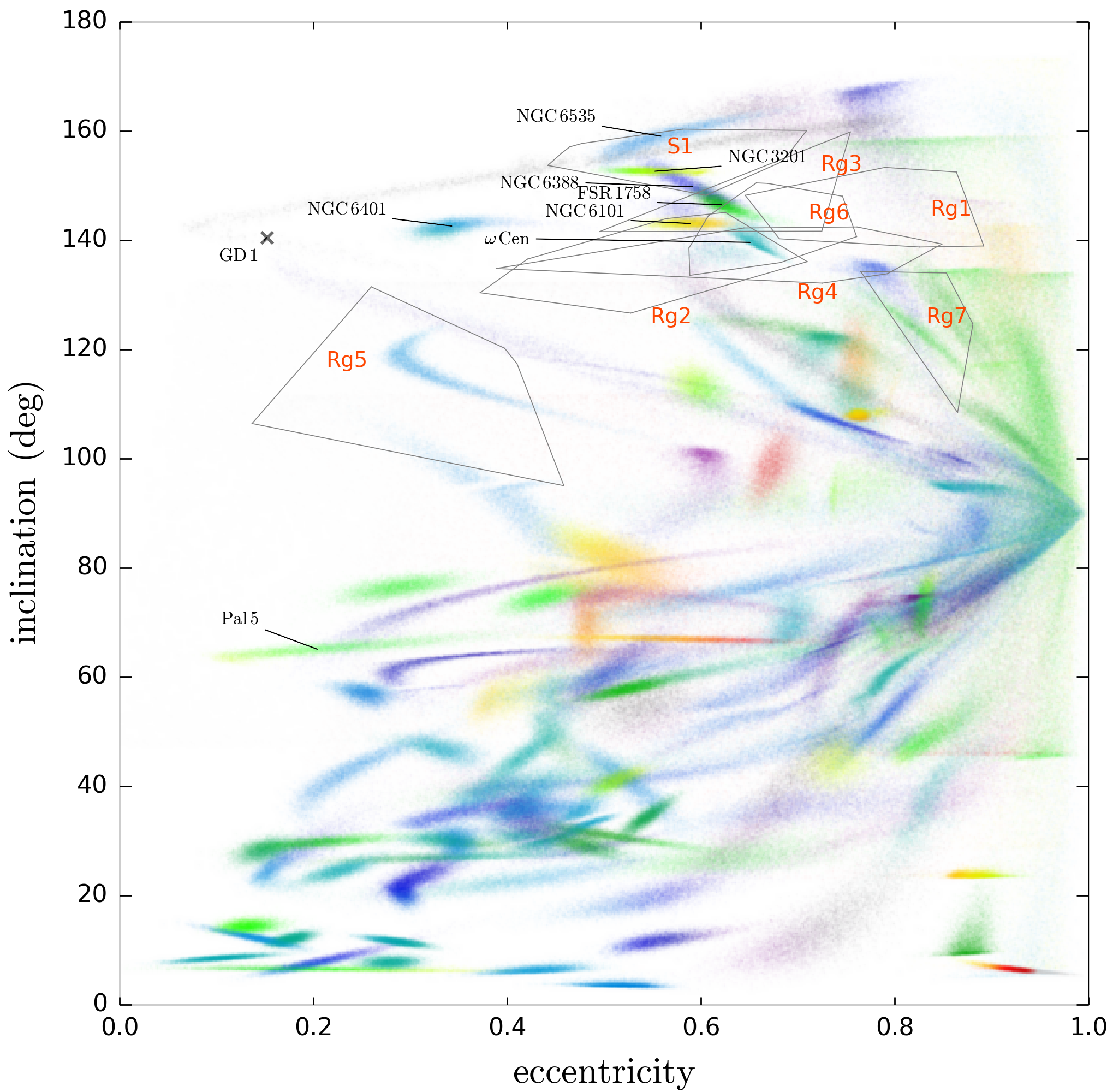}
 \end{center}
 \caption{Distribution of orbital eccentricity and inclination for the Milky Way GCs \citep{Vasiliev19} and retrograde substructures \citep{Myeong18_rg}. The GD-1 stream \citep{Grillmair06,Webb19} is marked with a cross. Colours have the same meaning as in Fig.~\ref{fig:actionmap}.}
 \label{fig:eccincl}
 \end{figure}
\begin{figure}
 \begin{center}
\includegraphics[width=0.45\textwidth]{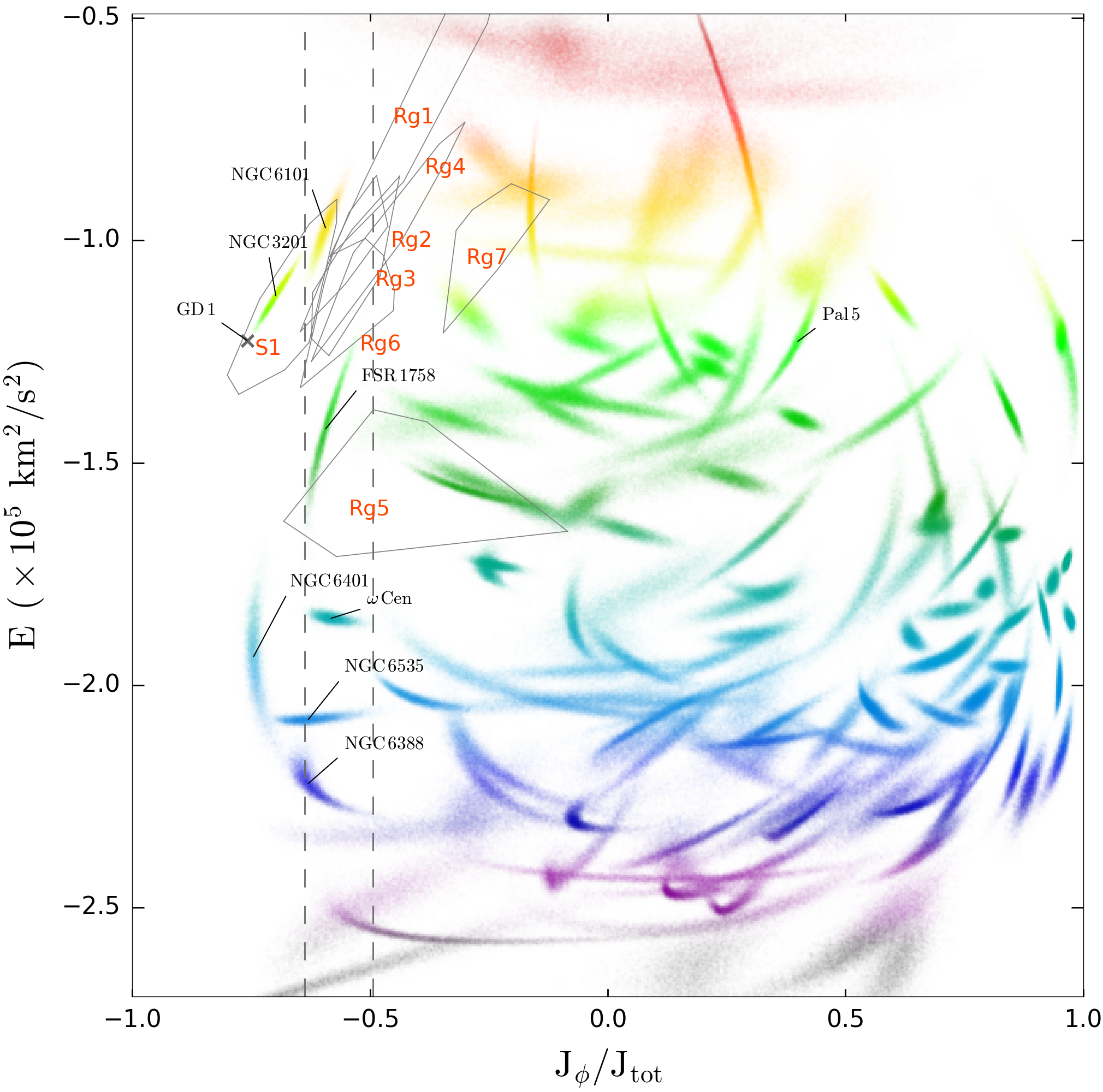}
 \end{center}
 \caption{Distribution of energy and normalised azimuthal action $J_{\phi} / J_{\mathrm{tot}}$
 for the Milky Way GCs \citep{Vasiliev19} and retrograde substructures \citep{Myeong18_rg}. The GD-1 stream \citep{Grillmair06,Webb19} is marked with a cross. Grey dashed lines are marking the $3\sigma$ range of $\omega$~Centauri (NGC~5139) in $J_{\phi} / J_{\mathrm{tot}}$, which traces the orbital inclination roughly. Candidate members associated with the accretion event that included $\omega$~Centauri are expected to lie within this range.}
 \label{fig:energyaction}
 \end{figure}

\section{Tracers of the Sequoia Galaxy}

FSR~1758 has a very retrograde, eccentric orbit \citep[e.g.,][]{Simpson19}. An accretion origin for the strongly retrograde components of the stellar halo has long been suspected~\citep[e.g.,][]{Quinn86, Norris89, Carollo07, Beers12}. Analyses of the \textit{Gaia} data \citep{Helmi17, Myeong18_ah, Myeong18_rg} have convincingly shown that the highest energy stars in the halo are typically retrograde, reinvigorating these earlier suspicions. This rotational asymmetry could be the consequence of dynamical interaction between accreted satellites and the Milky Way. For example, \citet{Quinn86} and \citet{Norris89} show that retrograde accretion events -- especially with some inclination -- experience much less drag than prograde ones. So, the effect of many random infalls can produce an the overall rotational asymmetry at high energy. A sharp feature traced by the retrograde stellar substructure has already been seen in energy-action space with a specific metallicity range. This is especially obvious in the subpanels of metallicity $-1.9 < \mathrm{[Fe/H]}<-1.5$ in Figure.~2 of \citet{Myeong18_ah}. This is a clear signature of an individual event of retrograde accretion in the past, distinct from other, numerous, past accretions.

Fig.~\ref{fig:actionmap} shows the Milky Way GCs from \citet{Vasiliev19}, together with the substructures found by \citet{Myeong18_rg}, in the plot of scaled action. GCs on retrograde orbits, including FSR~1758, lie on the far left-hand side of the plot. Specifically, the horizontal axis is the (normalized) azimuthal action $J_{\phi} / J_{\mathrm{tot}}$), while the vertical axis is the (normalized) difference between the vertical and radial actions  ($(J_z - J_R) / J_{\mathrm{tot}}$). Colour represents the radius of the circular orbit with the same energy, and so gives an idea of the typical distances probed by an object. The observational uncertainties such as the distance, line-of-sight velocities and proper motions are Monte Carlo sampled, with error ellipse transforming in action space to distended shapes. The gravitational potential used to represent the Milky Way is the one recommended as the best amongst the suite studied by \citet{McMillan17}. It is an axisymmetric model with bulge, thin, thick and gaseous disks and an NFW halo.

The portion of the plot occupied by FSR~1758 (coloured green) overlaps with a number of GCs, in particular $\omega$~Centauri. Using a `Friends-of-Friends' clustering algorithm in this scaled action space, we identify 6 GCs that form an agglomeration. They are FSR~1758, NGC~3201, $\omega$~Centauri (NGC~5139), NGC~6101, NGC~5635, and NGC~6388. All 6 are listed in Table~\ref{tab:table_1}, which gives their actions, energies and orbital characteristics. Also shown on Fig.~\ref{fig:actionmap} are all the retrograde stellar substructures identified by \citet{Myeong18_rg}. These are depicted as irregularly shaped polygons that include all the stars believed to be members. Apart from Rg5 and Rg7, it is striking that all the retrograde substructures overlap with our group of GCs associated with FSR~1758. Fig.~\ref{fig:eccincl} shows the same objects, but now plotted in the plane of eccentricity and inclination. We see that the GCs listed in Table~\ref{tab:table_1} are restricted to a narrow range of inclinations ($140^\circ-160^\circ$) and eccentricities ($e\sim 0.6$). We remark that additionally NGC~6401 may be associated with the group, at least as judged by inclination. However, its eccentricity is somewhat less than the other members. As a possible member, the orbital properties of NGC~6401 are also listed in the lower part of Table~\ref{tab:table_1}. It is noteworthy that the inclination range of our group appears to be distinct from the orbital plane of the Magellanic system \citep[e.g.,][]{DOnghia09,Nichols11} or the plane of Milky Way satellites \citep{Kroupa05} known to be near perpendicular to the Galactic plane.

In addition to the GCs \citep{Vasiliev19} and retrograde stellar substructures \citep{Myeong18_rg}, a retrograde stellar stream, GD-1 \citep{Grillmair06} is also marked on Fig.~\ref{fig:actionmap}, \ref{fig:eccincl} and \ref{fig:energyaction} based on a representative six-dimensional phase space information from \citet{Webb19}. Interestingly, the orbital inclination, (normalized) azimuthal action and energy of GD-1 appear to be comparable to our group of GCs and retrograde substructures. But, its other action components and orbital eccentricity noticeably differ from our group. According to the complex morphology of GD-1 \citep[see e.g.,][]{PriceWhelan18,Malhan19}, there is a possibility that the current orbital characteristics of GD-1 may not be a good reflection of its past state or its progenitor or its parent dwarf galaxy. Nonetheless, it will require more detailed investigation to find out any potential connection between our group and GD-1.

There have been long-held suspicions over the intriguing and anomalous globular cluster, $\omega$~Centauri. Normal globular clusters often have multiple populations, but show homogeneous abundances in heavy elements (such as calcium and iron) together with variations in light elements (such as the oxygen-sodium anticorrelation).
However, $\omega$ Centauri hosts multiple stellar populations with different heavy element abundances enriched by supernovae, as well as spreads in the light elements~\citep[see e.g.,][]{Lee99,Be04,DAn11, DO11,Joo13}. This necessitates the existence of different channels for enrichment to account for the chemical peculiarities of $\omega$~Centauri~\citep[e.g.,][]{Be06, Romano07, Marcolini07}. Already \citet{Fr93} and \citet{Be03} suggested it is the nucleus of a stripped dwarf galaxy, inspired by its very bound retrograde orbit. A number of authors have argued that the retrograde components in the stellar halo may be related to the disruption of $\omega$~Centauri \citep[e.g.,][]{Dinescu02, Brook03, Majewski12, Helmi17, Myeong18_ah, Myeong18_rg}. We amplify this hypothesis here by associating it with the Sequoia Event. Either $\omega$~Centauri is the remnant core of the Sequoia galaxy, or it was the largest GC member. For dwarf galaxies in cored haloes, the nucleus may be completely dissolved by the merging process, leaving only the GCs and stellar debris. For nucleated dwarfs or dwarfs in cusped haloes, the nucleus can survive intact, even if the outer parts are stripped. Whichever picture is correct, it is reasonable to conclude that the Sequoia dwarf once possessed an entourage of up to 7 GCs (5 probable and 2 possible).

\begin{figure}
 \begin{center}
\includegraphics[width=0.47\textwidth]{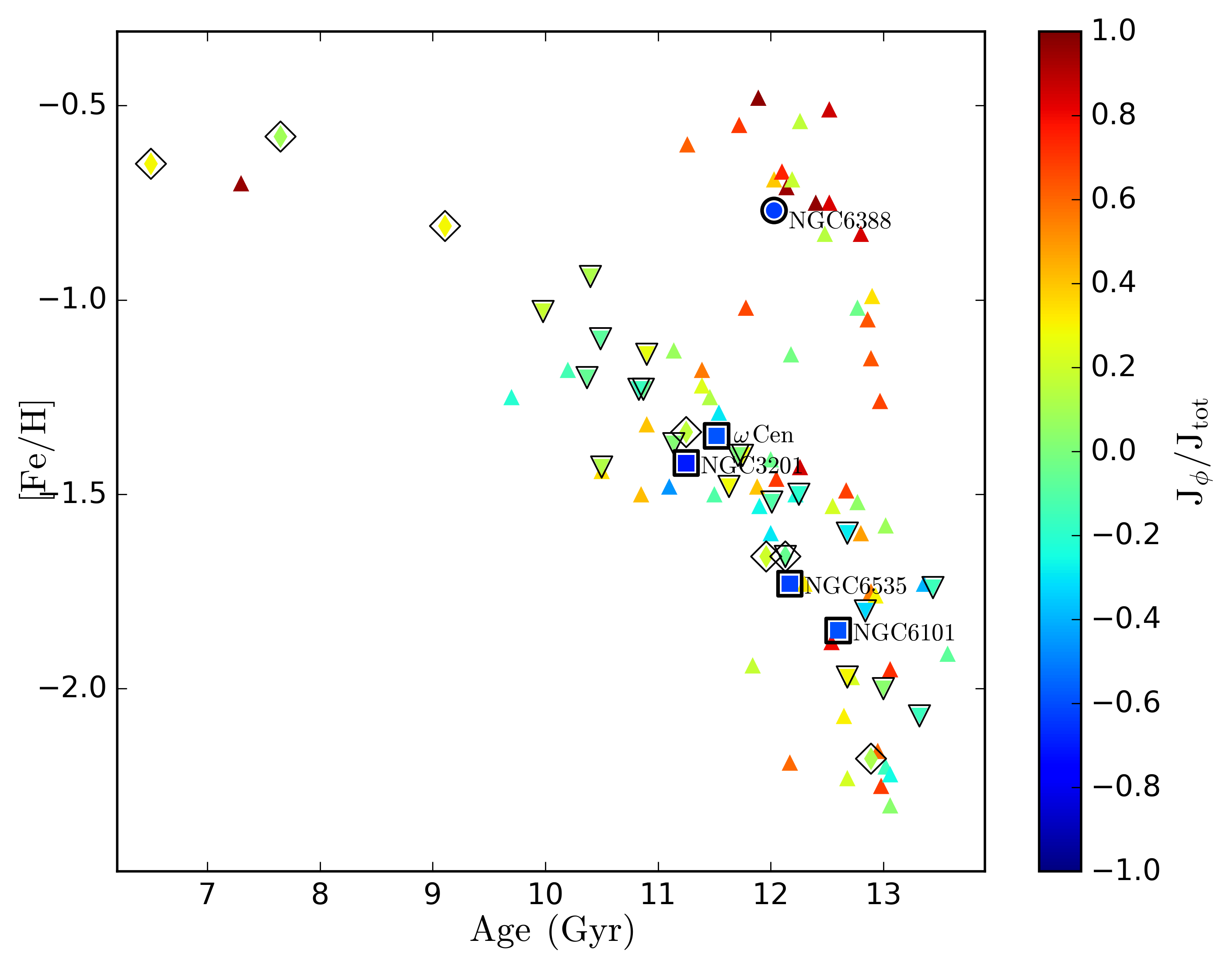}
\includegraphics[width=0.47\textwidth]{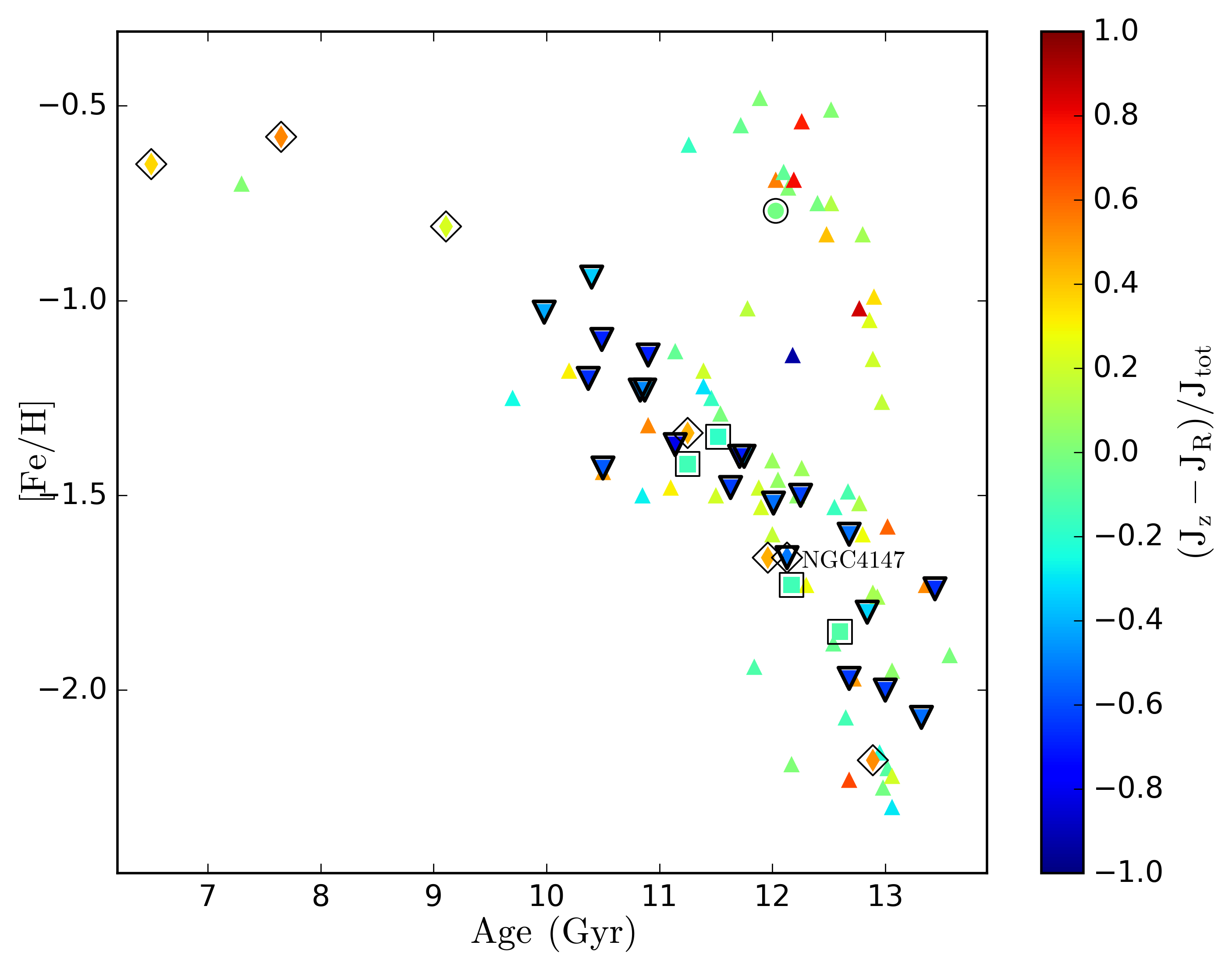}
 \end{center}
 \caption{Distribution of age and metallicity for the Milky Way GCs from \citet{Kr19} and the references therein. Five Sequoia member GCs with existing estimates are marked with squares (probable) and circle (possible). Sausage GCs \citep[see e.g.,][]{Myeong18_sg,Vasiliev19} are marked with downward-pointing triangles. Sgr GCs \citep[see e.g.,][]{Forbes10} are marked as diamonds. Upper panel: Colour shows the normalised azimuthal action $J_{\phi} / J_{\mathrm{tot}}$. The member GCs stand out clearly with $J_{\phi} / J_{\mathrm{tot}} \sim -0.6$. Lower panel: Colour shows the normalised difference between the vertical and radial actions $(J_z - J_R) / J_{\mathrm{tot}}$. Sausage GCs stand out clearly with $(J_z - J_R) / J_{\mathrm{tot}} \sim -0.6$.}
 \label{fig:agemetallicity}
 \end{figure}
Fig.~\ref{fig:energyaction} shows the distribution in energy and normalised azimuthal action. We show the azimuthal action of $\omega$~Centauri with its $3\sigma$ uncertainty as dashed vertical lines. If inclination is roughly preserved under the action of dynamical friction for these strongly retrograde mergers, then the dashed lines will delineate GCs and substructure associated with the $\omega$~Centauri, and hence the Sequoia. We see that FSR~1758, as well as the objects in Table~\ref{tab:table_1}, all lie within this band. GCs with higher energy were stripped earlier and/or composed the trailing tail. In this picture, FSR~1758 was one of the most massive GCs in the precursor dwarf galaxy, residing near the centre of the progenitor. So, it has ended up with comparable, if higher, energy. Objects with lower energy may have come from the disruption of the leading tail. This includes NGC~6535 (and possibly NGC~6388), which have been deposited closer to the centre of the Milky Way on tighter orbits.

From energy arguments, the retrograde substructures \citep{Myeong18_rg} are more likely to be the remnant debris of the Sequoia than direct tidal debris from $\omega$~Centauri. The retrograde stellar substructures are stripped well before the progenitor became severely destroyed. In this case, if its core is indeed $\omega$~Centauri, we do not expect the retrograde substructures to show the unique chemical abundance patterns observed from $\omega$~Centauri \citep[e.g., Na--O and Mg--Al patterns, Ba overabundance; cf.][]{Navarrete15}. These may have been imprinted on $\omega$~Centauri by subsequent accretion of gas onto the nucleus, after the stripping process removed the retrograde substructures. It would be interesting to revisit Kapteyn's Moving Group \citep{Wylie10} or the $\omega$~Centauri Moving group \citep{Meza05} -- which were disproved to be $\omega$~Centauri's tidal debris based on the chemical abundances \citep{Navarrete15} -- for the possibility that they are remnant debris of the Sequoia. By contrast, the freshly discovered Fimbulthul stream \citep{Ibata19} is probably stripped from $\omega$~Centauri itself rather than Sequoia and in this case should show the unique abundance patterns.

The age--metallicity relation of the GCs is shown in Fig.~\ref{fig:agemetallicity} based on data complied by \citet{Kr19}. Among five member GCs estimates four GCs (except NGC~6388) form a distinct track in the age--metallicity relation that is different from the Milky Way's in-situ GCs. This is most visible in the upper panel of Fig.~\ref{fig:agemetallicity}. This branching is similar to what has already been seen for GCs associated with major accretion events -- in particular, the Sagittarius (Sgr) GCs \citep[Terzan~7, Terzan~8, Arp~2, Pal~12, NGC~4147\footnote{As will be discussed later, this cluster is unlikely to belong to the Sgr group, based on its kinematics.}, NGC~6715, and Whiting~1,][and marked as diamonds]{Forbes10} or the Sausage GCs \citep[see e.g.,][and marked as downward-pointing triangles]{Myeong18_sg}. The track of the Sausage GCs stands out clearly especially at the lower panel of Fig.~\ref{fig:agemetallicity}.
%These other cases of ex situ GCs possess some vertical offset towards lower metallicity. 
This track in the age--metallicity plot of Fig.~\ref{fig:agemetallicity} is remarkable especially since the original membership of the GCs is established purely based on dynamical information. This is independent evidence of the extragalactic origin of the member GCs from an individual merger event. 

\begin{figure}
 \begin{center}
\includegraphics[width=0.47\textwidth]{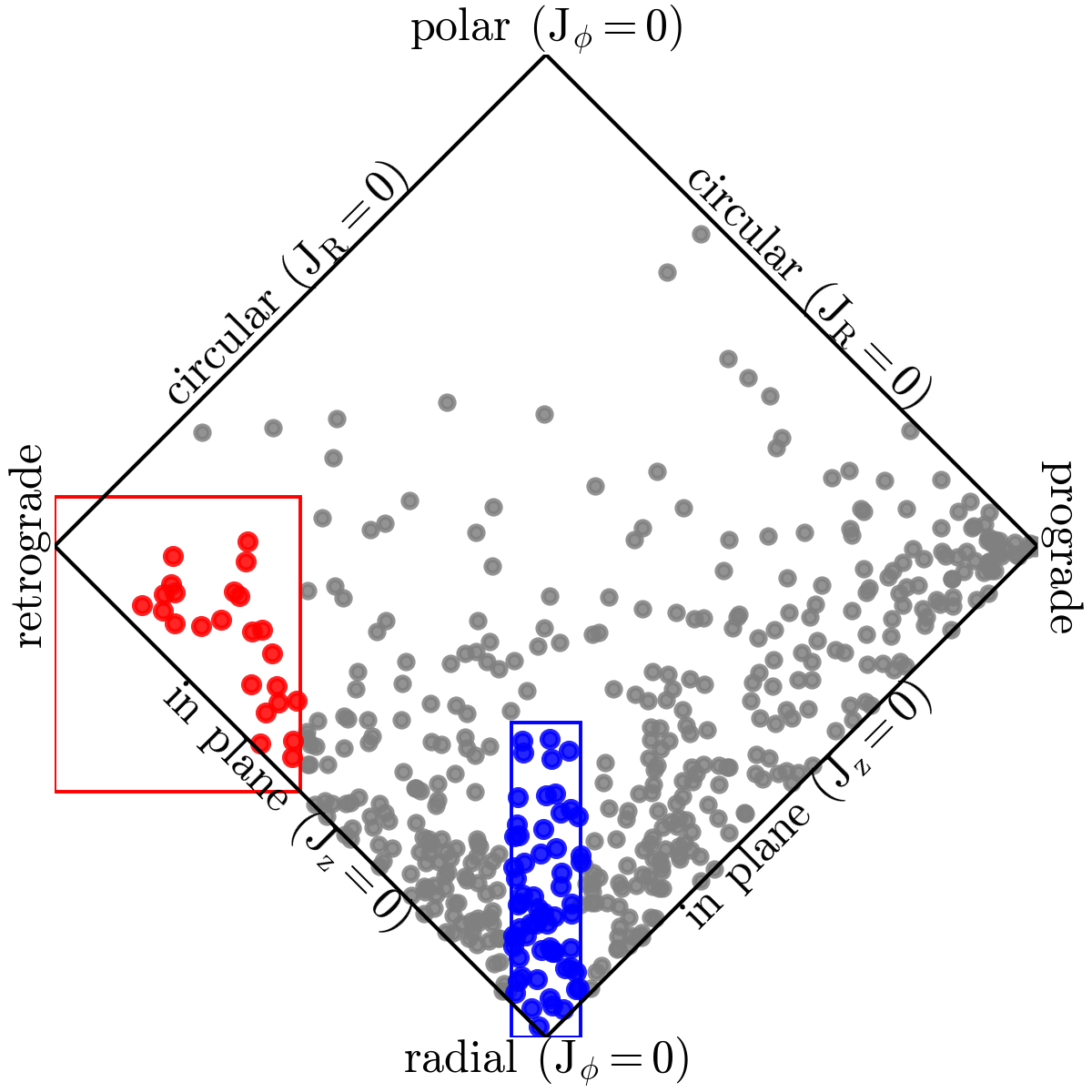}
\includegraphics[width=0.47\textwidth]{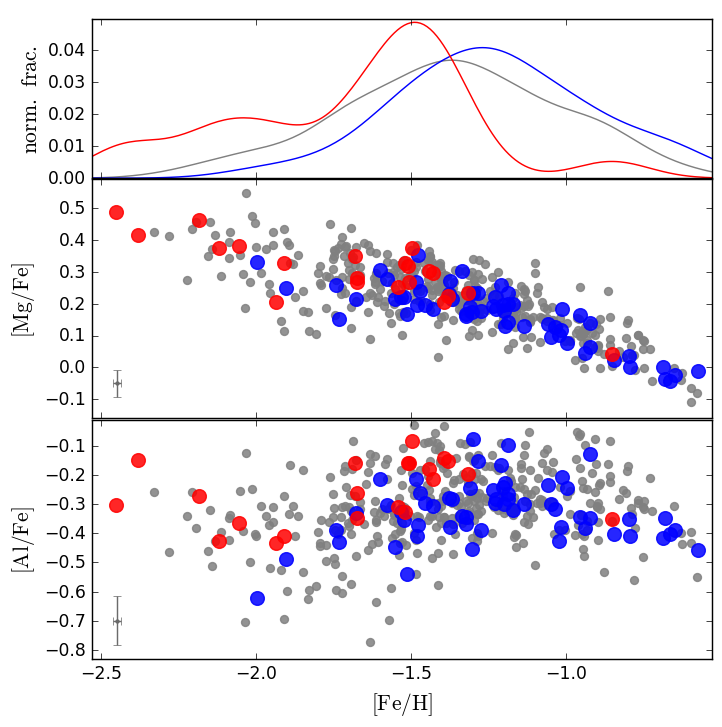}
 \end{center}
 \caption{The action-space map, metallicity distribution and abundance patterns for the halo stars in APOGEE DR14 \citep{Abolfathi18,Bovy19}. Gaia Sausage remnant set ($e\sim0.9$ with $\lvert J_{\phi} / J_{\mathrm{tot}} \rvert < 0.07$ and $(J_{z}-J_{R}) / J_{\mathrm{tot}} < -0.3$) and the high energy retrograde set ($e\sim0.6$ with $J_{\phi} / J_{\mathrm{tot}} < -0.5$ and $(J_{z}-J_{R}) / J_{\mathrm{tot}} < 0.1$) are shown with blue and red. Rest of the halo stars are shown in grey.}
 \label{fig:apogee_mg_al_feh}
 \end{figure}

Halo stars can be identified effectively based on chemical abundances such as [Al/Fe] and [Mg/Fe] \citep{Hawkins15}. With APOGEE DR14 \citep{Abolfathi18}, \cite{Ma19} showed that halo stars with high eccentricity orbits tend to have lower [Mg/Fe] on average compared to the rest of the halo stars. In Fig.~\ref{fig:apogee_mg_al_feh}, we show more specifically that the highly radial Gaia Sausage remnant stars ($e\sim0.9$ with $\lvert J_{\phi} / J_{\mathrm{tot}} \rvert < 0.07$ and $(J_{z}-J_{R}) / J_{\mathrm{tot}} < -0.3$) and the high energy retrograde substructure stars ($e\sim0.6$ with $J_{\phi} / J_{\mathrm{tot}} < -0.5$ and $(J_{z}-J_{R}) / J_{\mathrm{tot}} < 0.1$) have clearly different [Fe/H] distribution and different abundance patterns. The Sausage remnants show a metallicity distribution function peak at [Fe/H]$=-1.3$, whereas the high energy retrograde stars are more metal poor, with the peak at [Fe/H]$=-1.6$ \citep[both in good agreement with][]{Myeong18_ah,Matsuno19}. 
While the metallicitiy distributions of the Sausage and Sequoia stars overlap, at fixed [Fe/H], the two galaxies show distinct patters in the abundance of alpha elements. For example, at [Fe/H]$\sim-1.5$, the Sequoia debris are clearly more enhanced in Al compared to the Sausage.
Such differences in the abundances provide additional evidence that the accretion event that made the high energy retrograde stars in the halo is also chemically different from the Sausage event. Interestingly, their chemical characteristics -- the Sausage having higher [Fe/H] and lower abundance ratios compared to the Sequoia progenitor -- are in line with the trend observed by \citet{Ma19} from the EAGLE simulations \citep{Schaye15}. 

\cite{Matsuno19} searched through the SAGA database, which contains $\sim 880$ metal-poor stars with [Fe/H] $< -0.7$. They also found that the high energy retrograde stars are clearly distinct from the stars of the Sausage which dominates the inner halo~\citep{Belokurov18}. 
They reported that the `knee' in the abundance and metallicity plane differs by about 0.5 dex (at [Fe/H]$\sim-2$ for the Sausage and $\sim-2.5$ for the retrograde stars) which is another indication of their different origin. 

Our hypothesis is distinct from the proposal of \citet{He18}, who made a broad selection based on the azimuthal action $-1500$~kpc~\kms${} < J_\phi < 150$~kpc~\kms only, as opposed to using additional integrals of motion. \citet{He18} used this sample to suggest the \lq Gaia-Enceladus' accretion event. In our picture, this sample contains stars belonging to both the Gaia Sausage and the Sequoia. Our hypothesis is more closely related to the work of \citet{Ma19}, who divided halo stars according to eccentricity and showed that the low and high eccentricity groups have different abundance ratios and probably different origin. For the low eccentricity group, \citet{Ma19} suggested they are likely to be a mixture of in situ halo stars and many smaller accreted materials -- which includes the Sequoia debris. \citet{Matsuno19} finding of a different `knee' is important corroboratory evidence of this as well. 
%though it actually shows the reversed abundance ratio (Sausage being more alpha-enhanced).

\begin{figure}
 \begin{center}
\includegraphics[width=0.47\textwidth]{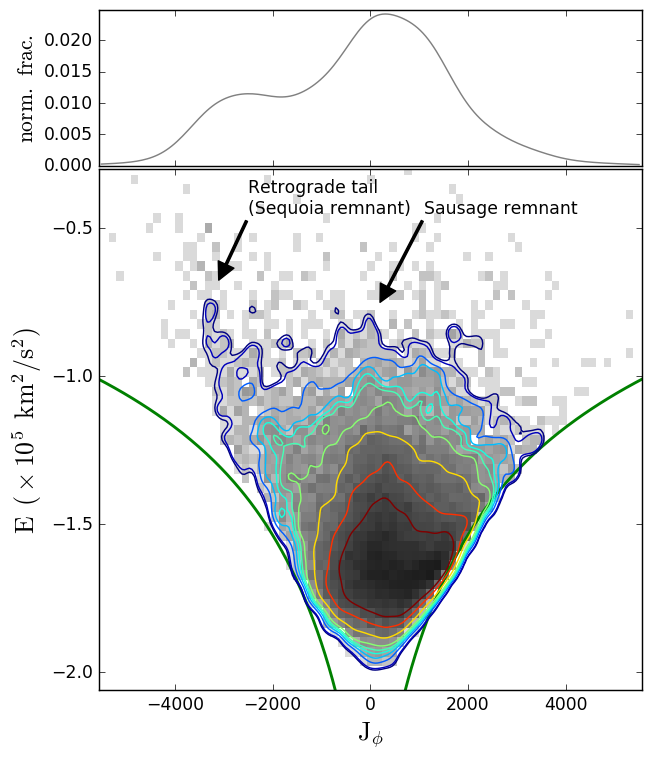}
 \end{center}
 \caption{Distribution of energy and azimuthal action for the halo stellar sample similar to Figure.~2 of  \citet{Myeong18_ah}. The top panel is showing the distribution function of the azimuthal action for the stars with high energy ($E>-1.1\times10^5$~km$^2$s$^{-2}$). The signature of Gaia Sausage remnants is visible as a peak at low $J_{\phi}$. A separate trace of a retrograde accretion (Sequoia event) is clearly visible as a sharp tail around $J_{\phi}\sim-3000$~kpc~\kms \citep{Myeong18_rg}. Green lines mark the circular orbit.}
 \label{fig:stellarJphiE}
 \end{figure}
\section{The Sequoia and the Sausage}

The progenitor of the Sausage had a total mass in stars and dark matter $\gtrsim 10^{11}$\msun \citep[][]{Belokurov18}. This was also derived in \citet{Fattahi19} using cosmological hydrodynamical simulations \citep{Grand17}, and is consistent with the estimates from \citet{Ma19} and \citet{Vincenzo19}. 
%The Sausage event is associated with perhaps as many as 12 GCs~\citep{Myeong18_sg,Vasiliev19}. 
\citet{Myeong18_sg} identified 10 GCs associated with the Sausage event from one of the earliest kinematics dataset of 75 Milky Way GCs based on the Gaia DR2 \citep{Gaia_Helmi18}. This Sausage GC membership has been revised with the more complete catalogue of 150 Milky Way GCs \citep{Vasiliev19} and also with the age--metallicity relation (see e.g., lower panel of Fig.~\ref{fig:agemetallicity}). The original identification by \citet{Myeong18_sg} was also limited to GCs with noticeably high energy only. These authors used the total energy of the young halo (YH) clusters~\citep[see e.g.,][for more details on the GC classification]{Zinn93,Mackey04,Mackey05} known at that time as a reference energy in order to concentrate on the old halo (OH) GCs with the most certain ex situ origin, judging from their noticeably high total energy. As we now have better knowledge on the typical chemo-dynamical characteristics of the Sausage GCs (e.g., $e\sim0.9$, $(J_z-J_R)/J_{tot}\sim-0.6$ with age--metallicity branching from Fig. 5 of \citet{Vasiliev19} and Fig.~\ref{fig:agemetallicity} in this work), we can revise the search in the larger catalogue of Milky Way GCs, without any artificial energy cut. We note that only the GCs with full kinematics, age and metallicity information have been examined. This gives up to 21 potential GCs showing typical chemo-dynamical characteristics of the Sausage, specifically NGC~362~\footnote{YH clusters~\citep[see e.g.,][]{Zinn93,Mackey04,Mackey05}\label{ftn:note}}, 1261~$^{\ref{ftn:note}}$, 1851, 1904, 2298, 2808, 4147~$^{\ref{ftn:note}}$\footnote{Note that NGC~4147 is among the potential Sausage GCs. NGC~4147 has previously been suggested as a Sgr GC \citep[e.g.,][]{Bellazzini03,Forbes10}, while there are studies suggesting no connection with Sgr dwarf \citep[e.g.,][]{Law10}. Here, we consider NGC~4147 to be a potential Sausage GC as the orbital characteristics of NGC~4147 are very different from the Sgr dwarf and other Sgr GCs, while they are similar to other Sausage GCs.}, 4833, 5286, 5694, 6544, 6584~$^{\ref{ftn:note}}$, 6712, 6779, 6864, 6934~$^{\ref{ftn:note}}$, 6981~$^{\ref{ftn:note}}$, 7006~$^{\ref{ftn:note}}$, 7089, Pal~14~$^{\ref{ftn:note}}$, Pal~15. 
Note that there are 8 YH classified GCs among the set with red horizontal branch (HB) morphology~\citep[see e.g.,][]{Zinn93,Mackey04,Mackey05}. Since the Sausage progenitor was considerably massive, its original GCs might have HB morphology similar to the Milky Way OH clusters~\citep[see][for more detail]{Mackey04,Mackey05}, while some of the younger GCs with red HB morphology might have been formed during the wet merger of the Sausage and the Milky Way~\citep[see e.g.,][]{Renaud18}, or have been acquired by the Sausage from separate accretion events before it merged with the Milky Way. Interestingly, this number, 13, of Sausage GCs with OH classification is in agreement with the suggested number of expected GCs originally formed in a major merger satellite of the Milky Way \citep[see e.g.,][]{Kr19}.

The Sequoia event has at least 4 GCs, excluding $\omega$~Centauri (as is right if it is the stripped core). The total stellar mass of GCs is known to be correlated with the (halo) mass of the host, albeit with some scatter. Based on the current and the initial masses of the GCs from~\citet{Baumgardt19} and using the fractional mass $\langle \eta \rangle = M_{\mathrm{GCs}} / M_{\mathrm{halo}} \sim 4\times10^{-5}$ derived by~\citet{Hudson14}, we can estimate the progenitor mass. In the case of the Gaia Sausage with 13 OH clusters, the progenitor mass is at least $1\times10^{11}$\msun (from the current mass of the GCs) or up to $4\times10^{11}$\msun (from the initial mass of the GCs). \citet{Hudson14} also provides a fractional value for the GC mass in terms of the stellar halo mass, which yields corresponding lower bounds on the stellar mass of the Sausage progenitor of $5\times10^{8}$\msun or up to $5\times10^{9}$\msun respectively. The abundance matching relation of \citet{Garrison14} gives similar estimates as well. Since our estimates are not redshift corrected, we note that they are overestimates of the actual progenitor mass at the merger time.

In comparison, the mass of the Sequoia galaxy is at least $1\times10^{10}$\msun (from the current mass of the GCs) or $5\times10^{10}$\msun (from the initial mass\footnote{For the mass fraction $M_{\mathrm{current}}/M_{\mathrm{initial}}$ for FSR~1758, we used the mean mass fraction of the other member GCs.} of the GCs). \citet{Valcarce11} suggested the mass of the progenitor of $\omega$~Centauri may be as high as $10^{10}$\msun from chemical evolution modelling of its multiple populations which is in broad agreement. The lower bound of the stellar mass from abundance matching gives $5\times10^{6}$\msun (current mass of GCs) or $7\times10^{7}$\msun (initial mass), while the relation from \citet{Hudson14} gives higher stellar mass estimates by a factor of two. The mass-metallicity relation of \citet{Kirby13} with a metallicity of $-1.6$ gives a broadly consistent stellar mass of $2\times10^{7}$\msun, but if we take account of the redshift evolution of the relation, this could be larger \citep[see e.g.,][]{Ma16}. For example, for $z=1.3$, the relation gives $1.7\times10^{8}$\msun which is again comparable to other estimates.

Although less massive then the Gaia Sausage, the Sequoia was a notable accretion in the evolutionary history of the Milky Way. In terms of the mass, the Fornax dwarf spheroidal could be a rough representation of the Sequoia progenitor. The fact that the Fornax dSph hosts a comparable number of GCs is also a similarity. Among its six GCs, Fornax~4 has been considered for possible ex situ origin \citep[see e.g.,][]{Buonanno99, vandenBergh00}, leaving Fornax with five probable in situ GCs. This is comparable to our estimates of the current number of member GCs of the Sequoia galaxy.

It is noteworthy that NGC~3201 is one of the probable members of the Sequoia event. It has already been pointed out that NGC~3201 is potentially associated with the S1 stellar stream~\citep{Myeong18_sc,OHare18}. In fact, NGC~3201 and the S1 stream have almost identical actions and energy. \citet{Myeong18_sc} identified the S1 stream as a stellar remnant of an accreted dwarf, and inferred its progenitor mass to be in order of $\sim10^{10}$\msun based on the library of accretion events using minor merger N-body simulations \citep{Amorisco17}. Interestingly, in this study, we already derived the lower bound mass of the Sequoia to be in order of $\sim10^{10}$\msun as well, based on completely different methods. Such agreement from independent approaches adds credence to the authenticity of this likely the same event discovered from two separate studies.

Now, the infall time for the S1 stream's progenitor is $\gtrsim 9$~Gyr~\citep{Myeong18_sc}. If the S1 progenitor is indeed the Sequoia, this provides a consistent picture, as the youngest GC associated with the event (NGC~3201) is $\sim11$~Gyr. In fact, the S1 stream and NGC~3201 bracket the range of possible infall times between 9 and 11 Gyr. Interestingly, \citet{Marcolini07} suggested that $\omega$~Centauri is a remnant of a dwarf spheroidal galaxy accreted $\sim10$~Gyr ago based on their study with hydrodynamical and chemical modelling. This is in a good agreement with our range. Also, we note that this suggested age range (9 Gyr to 11 Gyr) agrees very well with the suggested infall time of the Gaia Sausage itself. There is a possibility that the Sausage and the Sequoia galaxies were accreted at a comparable epoch. This suggests that here may have been a global association between them  -- perhaps the Sequoia was a satellite galaxy of the Sausage? The currently observed higher angular momentum of the Sequoia debris could reflect its binary orbital velocity at the time of accretion. More generally, any small differences in the initial condition at the early stage of the infall can easily cause the current difference between the orbital characteristics of the Gaia Sausage (highly radial with very low azimuthal action) and the Sequoia event (clearly retrograde). 

\section{Conclusions}

The starting point of our investigation is an unusual object FSR~1758 \citep{Froebrich2007}. Its enigmatic nature was recently pointed out by ~\citet{Barba19}, who raised the question as to whether FSR~1758 is an unusually large globular cluster or a dwarf galaxy remnant. Using Gaia data, we derived its proper motion dispersion profile, which is strongly declining, and so we concur with \citet{Simpson19} that FSR~1758 is an accreted, retrograde globular cluster. Our modelling suggests that FSR~1758 has a half-light radius of $\sim 9$ pc and a baryonic mass of $\sim 2.5\times 10^5\,M_\odot$ with an uncertainty $\lesssim 20\%$. 

It is natural to look for other retrograde globular clusters with similar actions as FSR~1758, which may have fallen in to the Milky Way at the same merger. This led us to the identification of Sequoia Event, which was already conjectured from our studies of stellar substructures~\citep[e.g.,][]{Myeong18_ah,Myeong18_rg}. Other investigators before us~\citep[e.g.,][]{Quinn86, Norris89, Carollo07, Beers12,Majewski12} have concurred that the highly retrograde parts of the stellar halo are most likely accreted.

The Sequoia Event is distinct from other known accretions, particularly the Gaia Sausage \citep{Belokurov18,Myeong18_ah,Myeong18_sg}. It has been seen in three tracers. First, there are at least 6 globular clusters, packed in action space around FSR~1758. These include $\omega$~Centauri itself, the most massive of the Milky Way globular clusters. Based on its kinematics and its spread of stellar ages, metallicities and abundances, this has long been suggested as the stripped core of a dwarf galaxy remnant~\citep[e.g.,][]{Be03,DAn11}. It may therefore be the remnant of the Sequoia galaxy. However, it is also possible that $\omega$~Centauri may have been a core globular cluster of a now wholly destroyed progenitor. Whichever hypothesis is correct, it still remains the case that FSR~1758 is also one of the Sequoia's largest globular clusters. Of the agglomeration of 6 globular clusters, 4 have existing age and metallicity estimates \citep[see e.g.,][and the references therein]{Kr19}. They form a distinct track in the age--metallicity relation, different from the Milky Way's in situ globular clusters. Additionally, this track shows evidence of an offset from the track of the Sausage globular clusters~\citep{Myeong18_sc}, supporting the identification of a separate event.

Secondly, the Sequoia Event is discernible in the retrograde stellar substructures, which are clearly visible in energy and action space~\citep{Myeong18_rg}. They form a separate grouping from the bulk of the Gaia Sausage, which has close to zero net angular momentum. This is clearly shown in Fig.~\ref{fig:stellarJphiE}, where the morphology of the contours in the high energy region shows the pattern of bimodal accretion tracks. The distribution of the azimuthal action for the stars with high energy (e.g., $E>-1.1\times10^5$~km$^2$/s$^2$) shows the existence of this extra retrograde component, clearly separated from the Sausage at zero angular momentum. \citet{Myeong18_ah,Myeong18_rg} showed the signal of this extra component is concentrated at a specific range of metallicity ([Fe/H] $\sim-1.6$). Thus, the Sequoia Event is also distinct from the Gaia-Enceladus structure \citep{He18}, which appears to combine parts of the Gaia Sausage and the Sequoia.

Thirdly, the very metal poor stars that are retrograde also have a chemical signature in the abundance and metallicity plane that is distinct from both the Sausage and the overall halo. This was already
hinted at in \citet{Ma19}, who used APOGEE data release 14 \citep[DR14,][]{Abolfathi18} to demonstrate that the retrograde halo stars  have lower [Mg/Fe] compared to the rest of the halo. The argument was further substantiated by \citet{Matsuno19}, who found evidence that the knee in the abundance and metallicity plane occurs at different spots separated by 0.5 dex for the stars in the Sausage and in the retrograde component or Sequoia. We have provided further evidence from APOGEE DR14 that the metallicity distribution and the abundance patters of Sausage and Sequoia stars are different (see Fig.~\ref{fig:apogee_mg_al_feh}). The peak of the metallicity distribution function of the Sausage is higher with [Fe/H]${} =-1.3$ as compared to the Sequoia at [Fe/H]${}=-1.6$. The abundance ratios of Sausage stars are lower than Sequoia stars.

These three lines of evidence all argue for two different accretion events. A number of arguments (abundance matching, mass in globular clusters today) suggest that the Sausage progenitor had a total mass of $\sim1-5\times 10^{11}$ \msun, whilst the Sequoia had a mass of $\sim1-5\times 10^{10}$ \msun. In terms of stellar mass, the Sausage weighs in at $\sim5-50\times 10^8$ \msun, whilst the Sequoia is $\sim5-70\times 10^6$ \msun. Modern day analogues would be the Large Magellanic Cloud and Fornax dSph, respectively. The infall time is somewhat comparable, and so the two progenitors may have been a binary pair or association. 

Different tracers (e.g., globular clusters, retrograde stellar substructures) stripped from the progenitor at different times now occupy different portions of energy and action space. They are like stepping stones that enable us to recreate the history of the event and trace out the time evolution of the disruption of the progenitor. To carry out such a re-creation, we need to be certain which substructures can be definitely associated with the Sequoia event. Here, detailed studies of abundances of stars with high resolution spectroscopy can play a crucial role. The chemical signature of the Sequoia event is seemingly evident both in Fig.~\ref{fig:apogee_mg_al_feh}, as well as in \citet{Ma19} and \citet{Matsuno19}. The current limitation is the size or the quality of the sample and so dedicated medium or high resolution spectroscopic follow-up study is essential.

\section*{acknowledgements}
We thank the referee for valuable comments and suggestions, which helped to improve the quality of the publication. GCM thanks the Boustany Foundation, Cambridge Commonwealth, European \& International Trust and Isaac Newton Studentship for their support of his work. EV acknowledges support from the Science and Technology Facilities Council of the United Kingdom, whilst GI thanks the Royal Society for the award of a Newton Fellowship.
The research leading to these results has received partial support from the European Research Council under the European Union's Seventh Framework Programme (FP/2007-2013) / ERC Grant Agreement no. 308024.
This work has made use of data from the European Space Agency (ESA) mission {\it Gaia} (\url{https://www.cosmos.esa.int/gaia}), processed by the {\it Gaia} Data Processing and Analysis Consortium (DPAC, \url{https://www.cosmos.esa.int/web/gaia/dpac/consortium}). Funding for the DPAC has been provided by national institutions, in particular the institutions participating in the {\it Gaia} Multilateral Agreement.

%%%%%%%%%%%%%%%%%%%%%%%%%%%%%%%%%%%%%%%%%%%%%%%%%%


\bibliographystyle{mnras}
\bibliography{./biblio} % if your bibtex file is called example.bib

%%%%%%%%%%%%%%%%%%%%%%%%%%%%%%%%%%%%%%%%%%%%%%%%%%

% Don't change these lines
%\bsp	% typesetting comment
\label{lastpage}
\end{document}